\documentclass[a4paper,fleqn,usenatbib]{mnras}

\usepackage[T1]{fontenc}
\usepackage{ae,aecompl}

\usepackage{graphicx}	
\usepackage{amsmath}	
\usepackage{amssymb}	

\usepackage{amsfonts, mathrsfs, times, float, color, bm}
\usepackage{extarrows}
\usepackage{graphicx,epstopdf}
\usepackage{dcolumn}
\usepackage{bm}
\usepackage{exscale}
\usepackage{relsize}

\newcommand{\nn}{\nonumber}

\title[radially moving Schwarzschild frequency shift]{Gravitational frequency shift of light in equatorial plane of a radially moving Schwarzschild black hole}

\author[G. He \& W. Lin]{
G. He$^{1}$
and W. Lin$^{1}$\thanks{wl@swjtu.edu.cn}
\\
$^{1}$School of Physical Science and Technology, Southwest Jiaotong University, Chengdu 610031, China
}

\begin{document}
\label{firstpage}
\maketitle

\begin{abstract}
The kinematical effect induced by the transversal motion of a gravitational lens on the frequency shift of light has been investigated in detail, while the effect of the radial motion is thought to be much smaller than the transversal one and thus has usually been neglected. In this work, we find that the radial velocity effect on the frequency shift has the same order of magnitude as that of the transversal velocity effect, when the light emitter (or the receiver) is close to the gravitational lens with the distance between them being an impact parameter scale. The significant velocity effect is usually transient due to the motion of the gravitational lens relative to the light emitter or the receiver.
\end{abstract}




\section{Introduction}
Translational motion of a gravitational system makes the field time-dependent and the induced velocity effect affects the propagation of electromagnetic waves. The issue of gravitational frequency shift of light caused by a moving central body has been studied by several groups~\citep{1983Nature...302...315B,1986Nature...324...349G,1988SAL...14...461K,1993ApJ...415...459P,1999PRD...60...124002K,2003ApJ...586...731M,2004PRD...69...063001W,2010MNRAS...402...650K,2013ApJ...774...70M,2015MNRAS...450...3155B}.
Especially, \citet{1983Nature...302...315B} investigated the transversal velocity effect of cluster of galaxies via a characteristic two-sided pattern of the brightness of an isotropic background radiation. They showed that the gravitational frequency shift of light is proportional to the transversal velocity of the gravitational lens in the low-velocity limit and the light deflection angle, which was further considered in the fully relativistic framework~\citep{1993ApJ...415...459P}.
Later, \citet{1999PRD...60...124002K} constructed their theory for the light propagation in the gravitational field of an arbitrarily moving N-body system, in which the frequency shift of light was calculated in the first post-Minkowskian (PM) approximation. \citet{2004PRD...69...063001W} studied the kinematical effect on the leading-order gravitational deflection of test particles and the leading-order frequency shift of light induced by a moving Schwarzschild source.

The previous work shows that the kinematical effect on the gravitational frequency shift caused by the transversal motion of the
gravitational lens is dominant relative to that of the lens' radial motion. This conclusion is indeed valid for the scenario in which both the light emitter and the receiver are located far away from the gravitational lens. However, when the emitter or the receiver is close to the gravitational lens, this is not the case, and the radial velocity effect has a similar order of magnitude to that of the transversal velocity effect, though the effective time for it is short in most astrophysical situations. In this work,
we re-study the velocity effect on the first-order gravitational frequency shift of light propagating in the equatorial plane of a radially moving Schwarzschild black hole, based on the Lorentz boosting technique. In order to compare the magnitude of the radial velocity effect with that of the transversal velocity effect on the gravitational frequency shift, we also re-derive the explicit form of the first-order frequency shift induced by a transversally moving Schwarzschild source. We restrict our discussions in the weak-field, small-angle, and thin lens approximation. For simplicity, we assume that both the light emitter and the receiver are static in the rest frame of the background.

This paper is organized as follows. In Section~\ref{derivation}, we first review the weak-field metric of the moving Schwarzschild source and then calculate the gravitational frequency shift up to the 1PM order. Section~\ref{Velocity-Effect} presents the discussions of the velocity correctional effect on the gravitational shift of frequency. The summary is given in Section~\ref{Summary}.

In what follows, we use natural units in which $G = c = 1$.
 \\
 \\

\section{gravitational frequency shift caused by a radially moving Schwarzschild source} \label{derivation}

\subsection{The 1PM metric for a radially moving Schwarzschild black hole} \label{2PM-Metric}

Let $\{\bm{e}_1,~\bm{e}_2,~\bm{e}_3\}$ be the orthonormal basis of a three-dimensional Cartesian coordinate system. $(t,~x,~y,~z)$ and $(X_0,~X_1,~X_2,~X_3)$ denote the rest frame of the background and the comoving frame for the barycentre of the gravitational lens, respectively. Then, the 1PM harmonic metric for a moving Schwarzschild black hole with a constant radial velocity $\bm{v}=v\bm{e_1}$ can be written as~\citep{2004PRD...69...063001W,2015RA&A...15...646H}
\begin{eqnarray}
&&\hspace*{-55pt}g_{00}=-1+\frac{2(1+v^2)\gamma^2M}{R}+O(M^2)~,~~~~ \label{g00MKN} \\
&&\hspace*{-55pt}g_{0i}=-\frac{4v\,\gamma^2M}{R}\delta_{i1}+O(M^2)~,~~~~\label{g0iMKN} \\
&&\hspace*{-55pt}g_{ij}=\left(1+\frac{2M}{R}\right)\!\delta_{ij}+\!\frac{4v^2\gamma^2M}{R}\delta_{i1}\delta_{j1}\!+\!O(M^2)~,~~~~~\label{gijMKN}
\end{eqnarray}
where $i$ and $j$ take values among the set $\{1,~2,~3\}$, $\delta_{ij}$ denotes Kronecker delta, and $\gamma=(1-v^2)^{-\scriptstyle \frac{1}{2}}$ is the Lorentz factor. $M$ is the rest mass of the gravitational lens, and $\Phi\!\equiv\! -\frac{M}{R}~(R>>M)$ denotes the Newtonian gravitational potential, with $R^2=X_1^2+X_2^2+X_3^2$. Note that the coordinates $X_0,~X_1,~X_2$, and $X_3$ are related to $t,~x,~y$, and $z$ by the Lorentz transformation
\begin{eqnarray}
&&\hspace*{-172pt} X_0\equiv T=\gamma(t-vx)~,  \label{LT-t} \\
&&\hspace*{-172pt} X_1\equiv X=\gamma(x-vt)~,  \label{LT-x} \\
&&\hspace*{-172pt} X_2\equiv Y=y~,             \label{LT-y} \\
&&\hspace*{-172pt} X_3\equiv Z=z~.             \label{LT-z}
\end{eqnarray}

\subsection{First-order frequency shift due to the moving Schwarzschild black hole} \label{1PM-FS}
\begin{figure*}
\begin{center}
  \includegraphics[width=15cm]{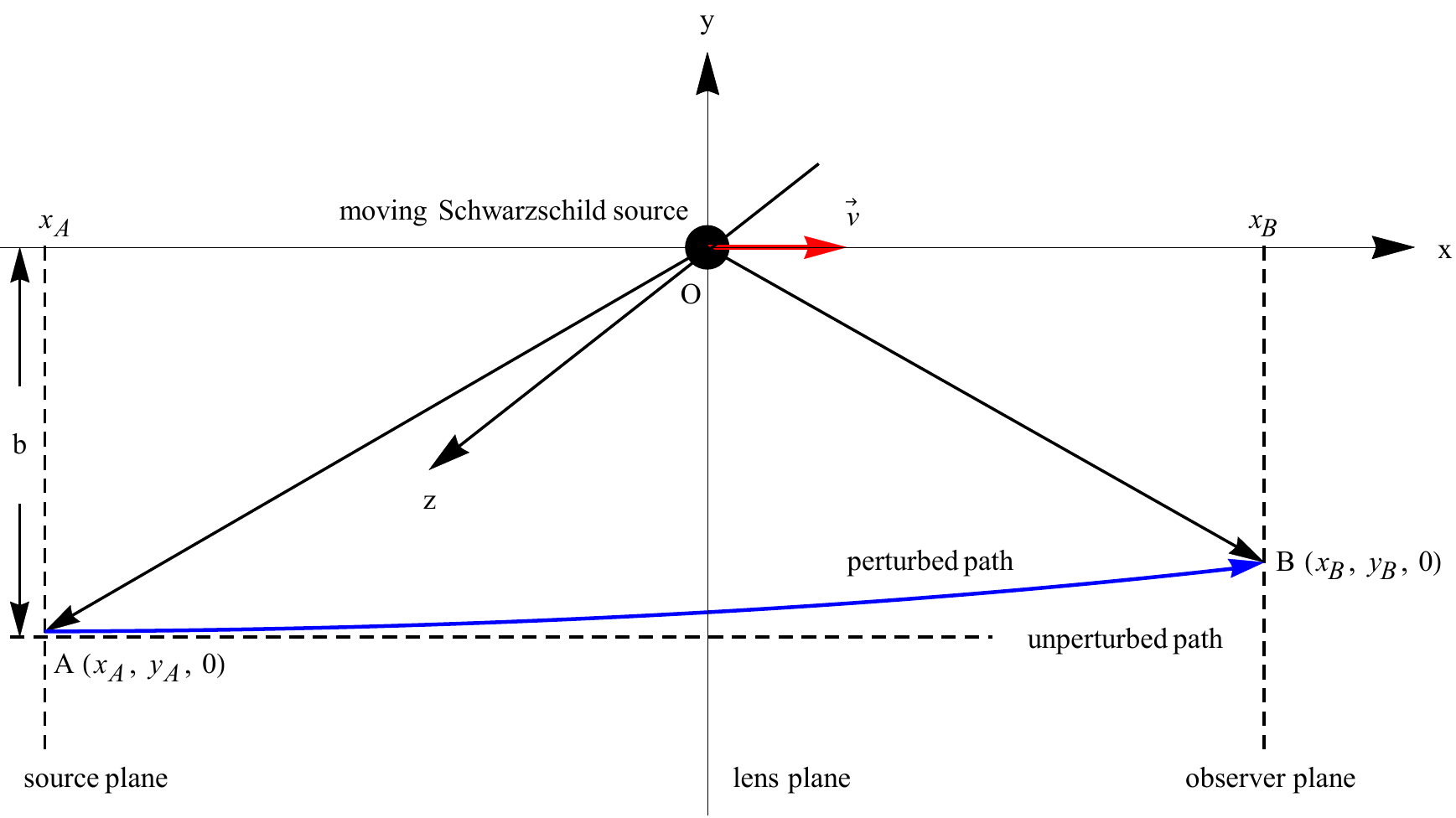}
  \caption{Schematic model for the light propagating in the gravitational field of the moving Schwarzschild black hole. The gravitational deflection is greatly exaggerated to distinguish the perturbed path (blue line) from the unperturbed path (dashed horizontal line). Notice that for the emission event $(t=t_A)$ the black hole is located at the position $(vt_A,~0,~0)$, while it is located at $(vt_B,~0,~0)$ for the reception event $(t=t_B)$. }    \label{Figure1}
\end{center}
\end{figure*}
We derive the gravitational frequency shift up to the 1PM order for the light propagating in the equatorial plane $(z=\frac{\partial }{\partial z}=0)$ of the moving Schwarzschild black hole. Figure \ref{Figure1} shows the schematic model for the propagation of light which is assumed to take the prograde motion relative to the rotation of the gravitational lens. The spatial coordinates of the light emitter (marked by $A$ on the source plane) and the receiver (marked by $B$ on the observer plane) are denoted as $(x_A,~y_A,~0)$ and $(x_B,~y_B,~0)$ in the background's rest frame, respectively, where $y_A<0$, $x_A<0$, and $x_B>0$. $b$ denotes the impact parameter~\citep{2016PRD...94...063011H}. The blue line represents the propagating path of a photon with its initial velocity $\bm{w}|_{x\rightarrow-\infty}~(=\bm{e}_1)$ being parallel to the black hole's velocity $\bm{v}$. The locations of $A$ and $B$ are denoted in the comoving frame by $(X_A,~Y_A,~0)$ and $(X_B,~Y_B,~0)$, respectively. Notice that we use $t_A$ and $t_B$ to denote the coordinate times of the emitter and the receiver, respectively.

The general form of the relative gravitational frequency shift is defined by~\citep{Synge1960,1999PRD...60...124002K}
\begin{equation}
\frac{\Delta \nu}{\nu}\equiv\frac{\nu_B}{\nu_A}-1\equiv \frac{d\tau_A}{d\tau_B}-1=\frac{d\tau_A}{dt_A}\frac{dt_A}{dt_B}\frac{dt_B}{d\tau_B}-1~, \label{FS1}
\end{equation}
where $\frac{d\tau_A}{dt_A}$ and $\frac{dt_B}{d\tau_B}$ for the light emitter $A$ and the receiver $B$ being static $(\bm{v}_A=\bm{v}_B=0)$ in the background's rest frame $(t,~x,~y,~z)$ are written as follows:
\begin{eqnarray}
&&\hspace*{-89pt}\frac{d\tau_A}{dt_A}=\sqrt{-g_{00}^A}
=\left[1\!-\!\frac{2\,(1+v^2)\gamma^2M}{R_A}\right]^{\frac{1}{2}}~, \label{FS2}   \\
&&\hspace*{-89pt}\frac{dt_B}{d\tau_B}=\frac{1}{\sqrt{-g_{00}^B}}
=\left[1\!-\!\frac{2(1+v^2)\gamma^2M}{R_B}\right]^{-\frac{1}{2}}~,  \label{FS3}
\end{eqnarray}
with $\tau_A$ and $\tau_B$ being the proper times of the light emitter and the receiver, respectively.

In order to calculate the analytical form of $\frac{dt_A}{dt_B}$, we need the 1PM time delay caused by the moving Schwarzschild black hole, which has been obtained in \citet{2016PRD...94...063011H} as
\begin{equation}
t_B-t_A=x_B-x_A+\Delta(t_B,~t_A)~, ~~ \label{2PMTimeDelay}
\end{equation}
with
\begin{eqnarray}
\hspace*{-8pt}\Delta(t_B,~t_A)=2(1\!-\!v)\gamma M\,ln\!\left(\!\frac{\sqrt{X_B^2+b^2}+X_B}{\sqrt{X_A^2+b^2}+X_A}\!\right)~. ~~~~~~~  \label{Delta}
\end{eqnarray}
Note that here $X_A=\gamma(x_A-vt_A)$ and $X_B=\gamma(x_B-vt_B)$.

For the case of $\bm{v}_A=\bm{v}_B=0$, the explicit form of $\frac{dt_A}{dt_B}$ can be calculated by~\citep{1999PRD...60...124002K}
\begin{equation}
\frac{dt_A}{dt_B}=\frac{1-\frac{\partial \Delta(t_B,~t_A)}{\partial t_B}}{1+\frac{\partial \Delta(t_B,~t_A)}{\partial t_A}}~,     \label{FS4}
\end{equation}
where $\frac{\partial \Delta(t_B,~t_A)}{\partial t_A}$ and $\frac{\partial \Delta(t_B,~t_A)}{\partial t_B}$ can be computed from Eq.~\eqref{Delta} as follows:
\begin{eqnarray}
&&\hspace*{-119pt}\frac{\partial \Delta(t_{\!B},\,t_{\!A})}{\partial t_A}=\frac{2vM}{(1+v)\sqrt{X_A^2\!+\!b^2}}~, \label{Delta-Partial-1}  \\
&&\hspace*{-119pt}\frac{\partial \Delta(t_{\!B},\,t_{\!A})}{\partial t_B}=-\frac{2vM}{(1+v)\sqrt{X_B^2\!+\!b^2}}~. \label{Delta-Partial-2}
\end{eqnarray}

In addition, the explicit form of $X_2$ up to the 0PM order can be written as~\citep{2016PRD...94...063011H}
\begin{equation}
X_2=y=-b+O(M)~, \label{y}
\end{equation}
which leads to the 0PM forms of $R_A$ and $R_B$ as follows:
\begin{eqnarray}
&&\hspace*{-152pt}R_A=\sqrt{X_A^2\!+\!b^2}+O(M) ~,    \label{RA}  \\
&&\hspace*{-152pt}R_B=\sqrt{X_B^2\!+\!b^2}+O(M) ~.    \label{RB}
\end{eqnarray}

Substituting Eqs.~\eqref{FS2} - \eqref{FS3} and \eqref{FS4} - \eqref{RB} into Eq.~\eqref{FS1}, up to the 1PM order, we can obtain the explicit form of the relative gravitational frequency shift caused by the moving Schwarzschild source
\begin{eqnarray}
\hspace*{-20pt}\frac{\Delta \nu}{\nu}=\left(1\!+\!2v\!-\!v^2\right)\gamma^2\!\left(\!\frac{M}{\sqrt{X_B^2\!+\!b^2}}\!-\!\frac{M}{\sqrt{X_A^2\!+\!b^2}}\right)~. \label{FS5}
\end{eqnarray}

In order to express Eq.~\eqref{FS5} in terms of the quantities in the background's rest frame $(t,~x,~y,~z)$, we adopt the 0PM forms of $X_A$ and $X_B$ given in \citet{2016PRD...94...063011H} as follows:
\begin{eqnarray}
&&\hspace*{-143pt}X_A=(1-v)\,\gamma\,x_A+O(M)~,    \label{XA}   \\
&&\hspace*{-143pt}X_B=(1-v)\gamma\,x_B+O(M)~.      \label{XB}
\end{eqnarray}
Plugging Eqs.~\eqref{XA} - \eqref{XB} into Eq.~\eqref{FS5}, we have
\begin{eqnarray}
\hspace*{-60pt}\frac{\Delta \nu}{\nu}=\left(1+2v-v^2\right)\!\gamma^2M\!\left(\frac{1}{S_B}-\frac{1}{S_A}\right)~,   \label{FS6}
\end{eqnarray}
with
\begin{eqnarray}
&&\hspace*{-142pt}S_A\equiv\sqrt{(1-v)^2\,\gamma^2\,x_A^2+b^2}~, \label{SA}  \\
&&\hspace*{-142pt}S_B\equiv\sqrt{(1-v)^2\,\gamma^2\,x_B^2+b^2}~. \label{SB}
\end{eqnarray}

It can be seen that for the non-moving Schwarzschild black hole ($v=0$), Eq.~\eqref{FS6} reduces to
\begin{eqnarray}
\hspace*{-68pt}\frac{\Delta \nu}{\nu}=M\!\left(\frac{1}{\sqrt{x_B^2+b^2}}\!-\!\frac{1}{\sqrt{x_A^2+b^2}}\right)~.   \label{FS7}
\end{eqnarray}

\section{Velocity effect on gravitational frequency shift} \label{Velocity-Effect}
In this section, we discuss the radial velocity effect on the gravitational frequency shift, and compare it with the transversal velocity effect for the leading-order frequency shift.

\subsection{Radial velocity effect on the gravitational frequency shift}
Comparing Eqs.~\eqref{FS6} and \eqref{FS7}, we can easily get the general form of the radial velocity effect on the first-order gravitational frequency shift as follows:
\begin{eqnarray}
\nn&&\hspace*{-35pt}\Delta_{M}(v)=\left(1+2v-v^2\right)\gamma^2M\left(\frac{1}{S_B}-\frac{1}{S_A}\right)-\frac{M}{\sqrt{x_B^2+b^2}} \\
&&\hspace*{5pt}+\frac{M}{\sqrt{x_A^2+b^2}}~. \label{Delta-FM}
\end{eqnarray}

\subsection{Transversal velocity effect on the first-order gravitational frequency shift}
In the literatures, the radial velocity effect is usually neglected in the investigation of the kinematical effect on the frequency shift. However, we find that the magnitude of the radial velocity effect may have a similar order to that of the transversal velocity effect. In order to compare them more conveniently, we re-derive the explicit form of the first-order gravitational frequency shift induced by a transversally moving Schwarzschild black hole.

For distinguishing the transversal comoving system from the radial one, we denote the comoving frame of the barycentre of a transversally moving gravitational lens as $(X_0^{'},~X_1^{'},~X_2^{'},~X_3^{'})$. Then, the 1PM harmonic metric for a moving Schwarzschild source along $y$-axis with a transversal velocity $\bm{v}=v_{T}\bm{e}_2$ can be written as~\citep{2015RA&A...15...646H}
\begin{eqnarray}
&&\hspace*{-49pt}g_{00}=-1+\frac{2(1+v_T^2)\gamma_T^2M}{R_T}+O(M^2)~,                                   ~~~~ \label{g00MS} \\
&&\hspace*{-49pt}g_{0i}=-\frac{4v_T\gamma_T^2M}{R_T}\delta_{2i}+O(M^2)~,                                         ~~~~ \label{g0iMS} \\
&&\hspace*{-49pt}g_{ij}=\left(1+\frac{2M}{R_T}\right)\delta_{ij}+\frac{4v_T^2\gamma_T^2M}{R_T}\delta_{2i}\delta_{2j}+O(M^2)~,~~~~ \label{gijMS}
\end{eqnarray}
where $\gamma_T=(1-v_T^2)^{-\scriptstyle \frac{1}{2}}$ denotes the Lorentz factor for the transversal motion and $R_T=\sqrt{X_1^{'2}+X_2^{'2}+X_3^{'2}}$. The coordinates $X_1^{'},~X_2^{'},~X_3^{'}$, and $X_0^{'}$ are related to $t,~x,~y$, and $z$ via the following Lorentz transformation
\begin{eqnarray}
&&\hspace*{-158pt} X_0^{'}\equiv T^{'}=\gamma_T(t-v_T y)~,  \label{LT-Transversal-t} \\
&&\hspace*{-158pt} X_1^{'}\equiv X^{'}=x~,                  \label{LT-Transversal-x} \\
&&\hspace*{-158pt} X_2^{'}\equiv Y^{'}=\gamma_T(y-v_T t)~,  \label{LT-Transversal-y} \\
&&\hspace*{-158pt} X_3^{'}\equiv Z^{'}=z~.                  \label{LT-Transversal-z}
\end{eqnarray}

Before we derive the gravitational frequency shift, we need to calculate the first-order time delay of the light propagating in the equatorial plane $(z=\frac{\partial }{\partial z}=0)$ of the moving Schwarzschild source. To do so, we need the explicit form of the coordinate time $t$ up to the 1PM order.
Following the same procedure in \citet{2016PRD...94...063011H}, we can obtain the integrating form of the 1PM coordinate time based on Eqs.~\eqref{g00MS} - \eqref{gijMS}
\begin{eqnarray}
\nn&&\hspace*{-59pt} t=\int\left(1+\frac{2\gamma_T^2M}{R_T} \right) dx+O(M^2)  \\
&&\hspace*{-53pt}=x+\int \frac{2\gamma_T^2M}{\sqrt{x^2+\gamma_T^2(y-v_T t)^2}} dx+C+O(M^2)~,     \label{t-1PM}
\end{eqnarray}
where $C$ denotes the integral constant which becomes $0$ if we choose a suitable value for $t_A$. For simplicity we choose $C=0$ here. Therefore, from Eq.~\eqref{t-1PM} we can get the zero$^{\text{th}}$-order form of $t$ as follow
\begin{equation}
t=x+O(M)~.  \label{t-0PM}
\end{equation}
Substituting Eq.~\eqref{t-0PM} into the second equality of Eq.~\eqref{t-1PM}, we have
\begin{eqnarray}
\nn&&\hspace*{-20pt} t\!=\!x\!+\!2\gamma_T M ln\!\left[x\!+\!v_T\gamma_T^2(v_T x\!-\!y)\!+\!\gamma_T\sqrt{x^2\!+\!\gamma_T^2(y\!-\!v_T x)^2}\right] \\
\nn&&\hspace*{-8pt} +\,O(M^2)  \\
&&\hspace*{-16pt}=x\!+\!2\gamma_T M ln\!\left(\!\gamma_T\sqrt{x^2\!+\!Y^{'2}}\!+\!x\!-\!v_T\gamma_TY^{'}\right)\!+\!O(M^2) ~.     \label{t-1PM-2}
\end{eqnarray}
From Eq.~\eqref{t-1PM-2}, we can obtain the 1PM form for the time delay of the light propagating in the equatorial plane of the moving lens from the emitter located at $(x_A,~y_A,~0)$ to the receiver located at $(x_B,~y_B,~0)$ as follow
\begin{equation}
t_B-t_A=x_B-x_A+\Delta(t_B,~t_A)_T+O(M^2)~,  \label{TimeDelay}
\end{equation}
where
\begin{equation}
\Delta(t_B,~t_A)_T= 2\,\gamma_T M\, ln\left(\!\frac{\gamma_T\sqrt{x_B^2\!+\!Y_B^{'2}}\!+\!x_B\!-\!v_T\gamma_TY_B^{'}}
{\gamma_T\sqrt{x_A^2\!+\!Y_A^{'2}}\!+\!x_A\!-\!v_T\gamma_TY_A^{'}}\!\right)~, \label{Delta-T}
\end{equation}
with
\begin{equation}
Y_A^{'}=\gamma_T(y_A-v_T t_A)~,~~~~~~Y_B^{'}=\gamma_T(y_B-v_T t_B)~.\label{YAp}
\end{equation}

From Eq.~\eqref{Delta-T} we have
\begin{eqnarray}
\nn&&\hspace*{-58pt}\frac{\partial \Delta(t_B,~t_A)_T}{\partial t_A}=\frac{\partial \Delta(t_B,~t_A)_T}{\partial Y_A^{'}}\frac{\partial Y_A^{'}}{\partial t_A}  \\
&&\hspace*{-2pt}=-\frac{2v_T\gamma_T^3M\left(v_T-\frac{Y_A^{'}}{\sqrt{x_A^2+Y_A^{'2}}}\right)}{\gamma_T\sqrt{x_A^2+Y_A^{'2}}+x_A-v_T\gamma_TY_A^{'}} ~,  \label{Partial-tA-T}  \\
\nn&&\hspace*{-58pt}\frac{\partial \Delta(t_B,~t_A)_T}{\partial t_B}=\frac{\partial \Delta(t_B,~t_A)_T}{\partial Y_B^{'}}\frac{\partial Y_B^{'}}{\partial t_B}  \\
&&\hspace*{-2pt}=\frac{2v_T\gamma_T^3M\left(v_T-\frac{Y_B^{'}}{\sqrt{x_B^2+Y_B^{'2}}}\right)}{\gamma_T\sqrt{x_B^2+Y_B^{'2}}+x_B-v_T\gamma_TY_B^{'}}~.  \label{Partial-tB-T}
\end{eqnarray}

Finally, substituting Eqs.~\eqref{g00MS}, \eqref{Partial-tA-T} and \eqref{Partial-tB-T} into Eq.~\eqref{FS1}, we can obtain the explicit form of the 1PM gravitational frequency shift caused by the transversally moving Schwarzschild black hole as follow
\begin{eqnarray}
\nn&&\hspace*{-23pt}\frac{\Delta\nu}{\nu}=\sqrt{\frac{1-\frac{2(1+v_T^2)\gamma_T^2M}{\sqrt{x_A^2+Y_A^{'2}}}}{1-\frac{2(1+v_T^2)\gamma_T^2M}{\sqrt{x_B^2+Y_B^{'2}}}}}
\frac{1-\frac{2v_T\gamma_T^3M\left(v_T-\frac{Y_B^{'}}{\sqrt{x_B^2+Y_B^{'2}}}\right)}{\gamma_T\sqrt{x_B^2+Y_B^{'2}}+x_B-v_T\gamma_TY_B^{'}}}
{1-\frac{2v_T\gamma_T^3M\left(v_T-\frac{Y_A^{'}}{\sqrt{x_A^2+Y_A^{'2}}}\right)}{\gamma_T\sqrt{x_A^2+Y_A^{'2}}+x_A-v_T\gamma_TY_A^{'}}}-1
\end{eqnarray}
\begin{eqnarray}
\nn&&\hspace*{-8pt}=\frac{2v_T\gamma_T^3M\left(\!v_T\!-\!\frac{Y_A^{'}}{\sqrt{x_A^2+Y_A^{'2}}}\!\right)}{\gamma_T\sqrt{\!x_{\!A}^2\!+\!Y_A^{'2}}\!+\!x_{\!A}\!-\!v_T\gamma_TY_{\!A}^{'}}
+\frac{(1\!+\!v_T^2)\gamma_T^2M}{\sqrt{x_B^2+Y_B^{'2}}}  \\
\nn&&\hspace*{-4pt}-\frac{2v_T\gamma_T^3M\left(\!v_T\!-\!\frac{Y_B^{'}}{\sqrt{x_B^2+Y_B^{'2}}}\!\right)}{\gamma_T\sqrt{\!x_{\!B}^2\!+\!Y_B^{'2}}\!+\!x_{\!B}\!-\!v_T\gamma_TY_{\!B}^{'}}
-\frac{(1\!+\!v_T^2)\gamma_T^2M}{\sqrt{x_A^2+Y_A^{'2}}}  \\
\nn&&\hspace*{-8pt}=\frac{2v_T\gamma_TM\!\left(v_T-\frac{y_A-v_Tx_A}{\sqrt{x_A^2+y_A^2-2v_T x_A y_A}}\right)}{\sqrt{x_A^2+y_A^2-2\,v_T\,x_A\,y_A}+x_A-v_T\,y_A}  \\
\nn&&\hspace*{-4pt}-\frac{2v_T\gamma_TM\!\left(v_T-\frac{y_B-v_Tx_B}{\sqrt{x_B^2+y_B^2-2v_T x_B y_B}}\right)}{\sqrt{x_B^2+y_B^2-2\,v_T\,x_B\,y_B}+x_B-v_T\,y_B}  \\
&&\hspace*{-4pt}+\frac{(1+v_T^2)\,\gamma_T\,M}{\sqrt{x_B^2\!+\!y_B^2\!-\!2\,v_T\,x_B\,y_B}}
\!-\!\frac{(1+v_T^2)\,\gamma_T\,M}{\sqrt{x_A^2\!+\!y_A^2\!-\!2\,v_T\,x_A\,y_A}}~, ~~~~~~   \label{FS32}
\end{eqnarray}
where we have made use of the zero$^{\text{th}}$-order relations $Y_A^{'}=\gamma_T(y_A-v_T x_A)+O(M)$ and $Y_B^{'}=\gamma_T(y_B-v_T x_B)+O(M)$.

It can be seen that for the case of both the light emitter and the receiver being far away from the moving gravitational lens with a low velocity ($x_A\ll -b,~x_B\gg b,~|v_T|\ll1$), up to the first order in velocity (FOV), Eq.~\eqref{FS32} reduces to~\citep{1983Nature...302...315B,1993ApJ...415...459P,1999PRD...60...124002K,2004PRD...69...063001W,2010MNRAS...402...650K}
\begin{equation}
\frac{\Delta\nu}{\nu}=\frac{2v_T\gamma_T M\left(v_T-\frac{y_A-v_Tx_A}{\sqrt{x_A^2+y_A^2-2v_Tx_Ay_A}}\right)}{\sqrt{x_A^2+y_A^2-2v_Tx_Ay_A}+x_A-v_Ty_A}
=\frac{4v_TM}{b}~. \label{FS33}
\end{equation}
Here, $y_A=-b+O(M)$ and $y_B=-b+O(M)$ have been used, and the terms with the factor $b^i/x_A^j$ or $b^i/x_B^j~(i=0,~1,~2,~...;~j=1,~2,~3,...)$ have been dropped. Notice that in \citet{1993ApJ...415...459P} and \citet{2004PRD...69...063001W} the deflection angle is negative since $y_A>0$ is chosen there.

Comparing Eq.~\eqref{FS7} with Eq.~\eqref{FS32}, we can write down the transversal velocity effect on the first-order gravitational frequency shift as
\begin{eqnarray}
\nn&&\hspace*{-34pt}\Delta_{M}(v_T)=\frac{2v_T\gamma_TM\!\left(v_T-\frac{y_A-v_Tx_A}{\sqrt{x_A^2+y_A^2-2v_T x_A y_A}}\right)}{\sqrt{x_A^2+y_A^2-2\,v_T\,x_A\,y_A}+x_A-v_T\,y_A}  \\
\nn&&\hspace*{-10pt}-\frac{2v_T\gamma_TM\!\left(v_T-\frac{y_B-v_Tx_B}{\sqrt{x_B^2+y_B^2-2v_T x_B y_B}}\right)}{\sqrt{x_B^2+y_B^2-2\,v_T\,x_B\,y_B}+x_B-v_T\,y_B}  \\
\nn&&\hspace*{-10pt}+\frac{(1+v_T^2)\,\gamma_T\,M}{\sqrt{x_B^2\!+\!y_B^2\!-\!2\,v_T\,x_B\,y_B}}
\!-\!\frac{(1+v_T^2)\,\gamma_T\,M}{\sqrt{x_A^2\!+\!y_A^2\!-\!2\,v_T\,x_A\,y_A}}  \\
&&\hspace*{-10pt}-\frac{M}{\sqrt{x_B^2+y_B^2}}+\frac{M}{\sqrt{x_A^2+y_A^2}}~.      \label{Delta-FM-T}
\end{eqnarray}

\subsection{Discussions}
We now concentrate on analyzing the magnitudes of the radial and transversal velocity effect on the frequency shift given in Eqs.~\eqref{Delta-FM} and \eqref{Delta-FM-T}, respectively. Throughout this subsection, the zero$^{\text{th}}$-order relations $y_A=-b+O(M)$ and $y_B=-b+O(M)$ will be used. For simplicity, our consideration is also limited in the FOV approximation (the low-velocity case) and we assume $|v|=|v_T|$.

\subsubsection{The case for $x_A\ll-b$ and $x_B\gg b$}
When both the light emitter and the receiver are far away from the gravitational lens, the magnitude of the transversal velocity effect is $|\Delta_{M}(v_T)|\approx\frac{4|v_T| M}{b}$, which is much lager than that of the radial velocity effect $|\Delta_{M}(v)|\approx3|v|M |\frac{1}{x_B}\!+\!\frac{1}{x_A}|$~. In other words, we can ignore the radial velocity effect when compared with the transversal velocity effect on the first-order gravitational frequency shift~\citep{1993ApJ...415...459P,1999PRD...60...124002K,2004PRD...69...063001W}.

\subsubsection{The case for $x_A\sim-b$ and $x_B\gg b$}  \label{FM-2}
When the $x$ position of the light emitter has the same order of magnitude as impact parameter $b$ while the receiver is still far away from the black hole, Eqs.~\eqref{Delta-FM-T} and \eqref{Delta-FM} reduce to
\begin{eqnarray}
\nn&&\hspace*{-53pt}\Delta_{M}(v_T)= \frac{2v_TM}{b}\left(1-\frac{x_A}{\sqrt{x_A^2+b^2}}\right)+\frac{v_T M b\, x_A}{\left(x_A^2+b^2\right)^{\frac{3}{2}}}  \\
&&\hspace*{-0.4cm}~\sim O(\frac{v_T M}{b})~,   \label{Delta-FM-T-2}  \\
&&\hspace*{-53pt}\Delta_{M}(v)= -\frac{vM(3\,x_A^2+2\,b^2)}{\left(x_A^2+b^2\right)^{\frac{3}{2}}} ~\sim O(\frac{v M}{b})~,  \label{Delta-FM-2}
\end{eqnarray}
where we have dropped the terms with the factor $v_T/x_B^2$ or $v/x_B$ since these terms are much smaller than the residual.
Eqs.~\eqref{Delta-FM-T-2} and \eqref{Delta-FM-2} show that the magnitude of the radial velocity effect has the same order as that of the transversal velocity effect on the leading-order gravitational shift of frequency, in other words, the radial velocity effect cannot be neglected any more. For example, if we set $x_A=-5b,~b=1.0\times 10^{5}M,~v=v_{T}=0.0001$, then, $\Delta_{M}(v)$ and $\Delta_{M}(v_T)$ will be about $-0.58\times10^{-9}$ and $3.92\times10^{-9}$, respectively. Especially, if $x_A=0$, we find that $\Delta_{M}(v_T)=\frac{2v_TM}{b}$ and $\Delta_{M}(v)=-\frac{2vM}{b}$, i.e., they have the same magnitude.

\subsubsection{The case for $x_A\ll -b$ and $x_B\sim b$} \label{FM-3}
When the light emitter is far away from the gravitational lens while $x_B$ has the same order of magnitude as $b$, Eqs.~\eqref{Delta-FM-T} and \eqref{Delta-FM} can be simplified as
\begin{eqnarray}
\nn&&\hspace*{-53pt}\Delta_{M}(v_T)= \frac{2\,v_T M}{b}\left(1+\frac{x_B}{\sqrt{x_B^2+b^2}}\right)-\frac{v_T M b\, x_B}{(x_B^2+b^2)^{\frac{3}{2}}}  \\
&&\hspace*{-0.4cm} ~\sim O(\frac{v_T M}{b})~,  \label{Delta-FM-T-3}  \\
&&\hspace*{-53pt}\Delta_{M}(v)=\frac{vM(3\,x_B^2+2\,b^2)}{(x_B^2+b^2)^{\frac{3}{2}}} ~\sim O(\frac{v M}{b})~,  \label{Delta-FM-3}
\end{eqnarray}
where the terms with the factor $v_T/x_A^2$ or $v/x_A$ have been dropped. In this case, the magnitude of the radial velocity effect also has the same order as that of the transversal velocity effect. Notice that Eqs.~\eqref{Delta-FM-T-3} and \eqref{Delta-FM-3} can also be obtained from Eqs.~\eqref{Delta-FM-T-2} and \eqref{Delta-FM-2} respectively by the exchange $x_A \rightarrow -x_B$, because of the reversal symmetry of the light path.

\subsubsection{The case for $x_A\sim-b$ and $x_B\sim b$}
In this case ($|x_A|\neq |x_B|$), both the light emitter and the receiver are close to the gravitational lens, and Eqs.~\eqref{Delta-FM-T} and \eqref{Delta-FM} reduce to
\begin{eqnarray}
\nn&&\hspace*{-25pt}\Delta_{M}(v_T)= v_T M\left[\frac{x_B(2x_B^2+b^2)}{b(x_B^2+b^2)^{\frac{3}{2}}}-\frac{x_A(2x_A^2+b^2)}{b(x_A^2+b^2)^{\frac{3}{2}}}\right]   \\
&&\hspace*{0.7cm}\sim O(\frac{v_T M}{b})~,  \label{Delta-FM-T-4}  \\
&&\hspace*{-25pt}\Delta_{M}(v)= vM\!\left[\frac{3x_B^2+2b^2}{(x_B^2+b^2)^{\frac{3}{2}}}-\frac{3x_A^2+2b^2}{(x_A^2+b^2)^{\frac{3}{2}}}\right]
\sim O(\frac{v M}{b})~.~~~~~~ \label{Delta-FM-4}
\end{eqnarray}
Again, the magnitude of the radial velocity effect also has the same order as that of the transversal velocity effect in this case.

It should be pointed out that the radial velocity effect is only significant for a small time in most scenarios, because of the motion of the gravitational lens relative to the light source or the receiver.

\section{Summary} \label{Summary}
We have studied the velocity effect on the first-order gravitational frequency shift of light propagating in the equatorial plane of a radially moving Schwarzschild source. It is found that the contribution of the radial velocity effect to the gravitational frequency shift has the same order of magnitude as that of the transversal velocity effect and thus can not be neglected, when the light emitter or the receiver is close to the gravitational lens. The significant velocity effect is usually only transient.

\section*{ACKNOWLEDGEMENT}
We would like to thank the anonymous reviewer for his/her constructive comments and suggestions on improving the quality of this paper. This work was supported in part by the National Natural Science Foundation of China (Grant Nos. 11647314 and 11547311).









\bsp	
\label{lastpage}
\end{document}